\documentclass[sigplan,twocolumn]{acmart}

\makeatletter
\def\@ACM@checkaffil{
    \if@ACM@instpresent\else
    \ClassWarningNoLine{\@classname}{No institution present for an affiliation}
    \fi
    \if@ACM@citypresent\else
    \ClassWarningNoLine{\@classname}{No city present for an affiliation}
    \fi
    \if@ACM@countrypresent\else
        \ClassWarningNoLine{\@classname}{No country present for an affiliation}
    \fi
}
\makeatother

\renewcommand\footnotetextcopyrightpermission[1]{}

\usepackage{tikz}
\usepackage{amsmath}

\usepackage{url}
\usepackage{xspace}
\usepackage{enumitem}

\usepackage{subcaption}

\usepackage{multirow}

\usepackage{calc}

\usepackage{bbding}

\usepackage{tablefootnote}

\usepackage{xcolor}
\usepackage{listings}

\usepackage{pifont}

\usepackage{array}

\usepackage{booktabs}  
\usepackage{geometry}  
\usepackage{tabularx}  
\usepackage{amsmath}

\usepackage{breqn}

\newcommand{\sys}{DHelix\xspace}
\newcommand{\flows}{strands\xspace}
\newcommand{\flow}{strand\xspace}
\newcommand{\si}{Strand Interleaving\xspace}
\newcommand{\SI}{SI\xspace}
\newcommand{\afl}{$\alpha$-strand\xspace}
\newcommand{\bfl}{$\beta$-strand\xspace}

\newcommand{\local}{node-local\xspace}
\newcommand{\cross}{cross-node\xspace}

\newcommand{\para}[1]{\noindent \textbf{#1 }}

\newcommand{\ie}{\textit{i.e.}\xspace}
\newcommand{\eg}{\textit{e.g.}\xspace}

\begin{document}

\title{Hiding Communication Cost in Distributed LLM Training via Micro-batch Co-execution}

\sloppy

\author{Haiquan Wang$^\ast$}
\affiliation{%
  \institution{University of Science and \\Technology of China}
  }
\email{whq1516@mail.ustc.edu.cn}

\author{Chaoyi Ruan$^\ast$}
\affiliation{%
  \institution{University of Science and \\Technology of China, Mohamed bin Zayed University \\of Artificial Intelligence}
  }
\email{rcy@mail.ustc.edu.cn}

\author{Jia He}
\affiliation{%
  \institution{University of Science and \\Technology of China, Mohamed bin Zayed University \\of Artificial Intelligence}
  }
\email{hej148@mail.ustc.edu.cn}

\author{Jiaqi Ruan}
\affiliation{
  \institution{University of Science and \\Technology of China, Mohamed bin Zayed University \\of Artificial Intelligence}
  }
\email{jiaqiruan@mail.ustc.edu.cn}

\author{Chengjie Tang}
\affiliation{
  \institution{Shanxi University}
  }
\email{tangcj@sxu.edu.cn}

\author{Xiaosong Ma}
\affiliation{
  \institution{Mohamed bin Zayed University \\of Artificial Intelligence}
  }
\email{xiaosong.ma@mbzuai.ac.ae}

\author{Cheng Li}
\affiliation{
  \institution{University of Science and \\Technology of China}
  }
\email{chengli7@ustc.edu.cn}

\begin{abstract}
The growth of Large Language Models (LLMs) has necessitated large-scale distributed training. 
Highly optimized frameworks, however, still suffer significant losses in Model FLOPS utilization (often below 50\%) due to large communication volumes. 
Meanwhile, our comprehensive profiling shows that the computation- and communication-intensive operators overlap well. 

This paper introduces \sys, a novel micro-structure that dramatically improves the efficiency of LLM training inspired by the DNA structure. 
Central to \sys's design is \textit{Strand Interleaving (SI)}, which views the continuous stream of training micro-batches through a GPU as two \textit{\flows}. 
\sys juxtaposes the forward and backward passes of the two \flows and 
performs a systematic optimization for an \SI plan that co-schedules the operators from the opposite \flows, enabled by operator-level overlap profiling results and a dynamic-programming based search algorithm. 
Meanwhile, \sys enables the two \flows to share model states and space for activation data, effectively accommodating two micro-batches with under 3\% extra memory space.
\sys seamlessly integrates with \textit{all forms of existing data/model parallelism}, the most challenging being pipeline parallelism, thanks to its unique \textit{model folding} design that results in a W-shaped pipeline.

We evaluate \sys training with the popular Llama and GPT dense models, plus the Phi Mixture of Expert (MoE) model, across 3 GPU clusters (A40, A800, and H100). 
Results show that it achieves 12-40\% (up to 58\% MFU) and 2-29\% (up to 71\% MFU) improvement on the 64-A40 and 64-A800 clusters, respectively, significantly outperforming state-of-the-art methods.
On the H100 cluster, though the faster network reduces \sys's profit margin, it makes cross-node tensor parallelism promising, a practice currently 
prohibitive due to communication costs.
\end{abstract}

\thanks{
$^\ast$Haiquan Wang and Chaoyi Ruan equally contributed to this work. This work was done when Chaoyi Ruan, Jia He and Jiaqi Ruan were visiting students at MBZUAI.
}

\maketitle
\pagestyle{plain}

\section{Introduction}
\label{sec:intro}
Recent advancements in Generative AI, particularly in areas such as chatbots~\cite{bhayana2024chatbots} and text generation~\cite{ni2021t5,dubey2024llama,brown2020language}, have driven a significant trend in Large Language Model (LLM) training. These LLMs, exemplified by models like Llama~\cite{dubey2024llama} and GPT~\cite{brown2020language}, are predominantly based on transformer architectures. As model sizes escalate from billions to trillions of parameters, 
distributed model training has become indispensable. However, current training methods for LLMs face a critical challenge: overall GPU throughput remains suboptimal, with end-to-end Model FLOPS utilization (MFU) falling below 50\%~\cite{jiang2024megascale}, wasting precious GPU cluster resources. 

A primary limiting factor performance is the bottleneck created by \textit{intra-layer communication}. 
Such communication emerges from various parallelism strategies, including Tensor Parallelism (TP)~\cite{shoeybi2019megatron}, Sequence Parallelism (SP)~\cite{korthikanti2023reducing}, Context Parallelism (CP)~\cite{megatron-cp,liu2023ring}, and Expert Parallelism (EP)~\cite{he2021fastmoe,liu2024deepseek}. 
They are crucial for mitigating the rapidly increasing activation memory consumption associated with growing model sizes and input parameters (sequence length and batch size). While effectively distributing computational tasks across multiple devices, these strategies introduce numerous communication operators (e.g., \texttt{AllGather}~\cite{nccl-allgather} and \texttt{ReduceScatter}~\cite{nccl-reduce-scatter}) into the critical training execution path. 
As the LLM layer size grows, such communication consumes considerable portions of the total execution time (Figure~\ref{fig:comp_ic_ratio}) and delays subsequent computation.

Two main approaches have been proposed to address this issue. 
The first is ``intra-batch''~\cite{jiang2024megascale,chang2024flux,wang2022overlap,jangda2022breaking}, overlapping computation and communication within a single micro-batch by breaking down these operators into smaller units. However, it has two significant drawbacks: (1) limited overlapping potential due to data dependencies within a micro-batch, and (2) degraded computational efficiency due to splitting well-optimized computation operators.

The second, ``inter-batch''~\cite{wang2021wavelet}, explores the concurrent 
execution of two batches, leveraging their complementary GPU memory usage to overlap the forward computation of one batch with the backward of another. 
However, as to be detailed later in this paper, it has fundamental limitations that hampers its application to frameworks using pipeline parallelism (PP), a major inter-layer mechanism for scaling out LLM training today. 
In addition, it adopts simple round-robin scheduling between the two micro-batches and poses large memory requirement due to model replication.

In this research, we start by comprehensively analyzing the trends in intra-layer communication growth and conduct operator-level profiling to study their pairwise performance behavior when overlapped by co-scheduling. 
Our results reveal that though communication sizes continue to grow, there are plenty of opportunities so far unexplored by existing frameworks.

\begin{figure}[!htb]
 \centering
 \includegraphics[width=0.46\textwidth]{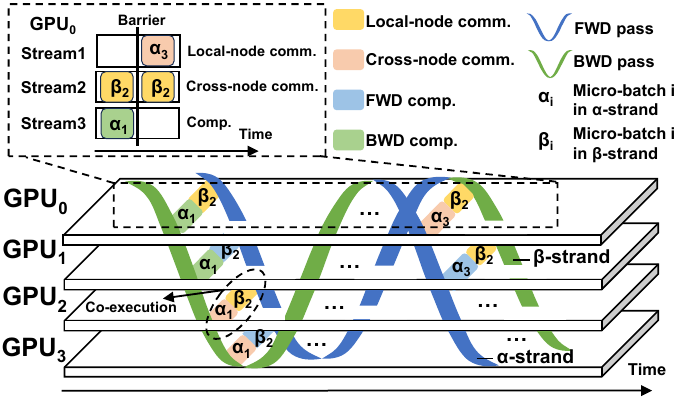}
 \caption{Double-\flow execution in \sys on 4 GPUs
 }
 \label{fig:overview}
\end{figure}

Based on these results and inspired by the DNA structure, we propose \sys (``Double-Helix LLM'' training),
a novel micro-structure that significantly improves the efficiency of LLM training. 
The main idea in \sys's design is \textit{Strand Interleaving (SI)}, which views the continuous stream of training micro-batches through a GPU as two \textit{\flows} to be co-executed, \afl and \bfl, as illustrated in Figure~\ref{fig:overview}.
\sys juxtaposes the forward (blue) and backward (green) passes of the two \flows together, which share model states and have complementary memory space usage for activation data. 

Within each GPU (top-left zoom-in illustration in Figure~\ref{fig:overview}), 
The operators that we profiled earlier now form a core set of \textit{operator bases}, computation-intensive or communication-intensive, to be ``paired'' with cross-\flow ``bonds'', like nucleotide bases on the DNA strands. 
\sys systematically searches for 
an optimal \SI plan that co-schedules the operators from the opposite \flows as allowed by the dependency within each \flow.
The search is based on its operator-level overlap profiling results and assisted by a dynamic-programming algorithm. 
Segments of these paired operators (colored blocks in Figure~\ref{fig:overview}) are co-scheduled by \sys-inserted cross-\flow barriers, with
their actual co-execution administered via three separate CUDA streams performing computation, local-node communication, and cross-node communication, respectively.

However, it is challenging to co-locate the two \flows going in opposite directions (forward and backward) in one GPU in the first place, with the prevalent pipeline parallelism (PP) adopted today.
To this end, we introduce a novel \textit{model folding} technique that transforms the double-\flow structure into an abstract U-shaped model.
Under PP, this folding technique leads to a W-shaped pipeline schedule that significantly reduces memory overhead by allowing both \flows to reuse the same parameter set on each GPU instead of requiring model replication.

Our comprehensive evaluation used  popular models such as dense model Llama, GPT as well as sparse model Phi.
Results demonstrate a significant improvement in overall training throughput across three NVIDIA GPU clusters: up to 40\% on a 64-card A40 cluster and up to 29\% on a 64-card A800 one, also considerably outperforming existing optimizations that we implemented following literature. 
On the H100 cluster, though the fast interconnection reduces our improvement margin, \sys makes cross-node tensor parallelism promising, a practice currently prohibitive due to communication costs.
Regarding memory efficiency, \sys supports a maximum model size of up to 97.5\% of the ideal Megatron-LM single-\flow limit, while providing up to 39\% performance improvement.

\section{Background}
\label{sec:backg}

\subsection{Distributed LLM training}
\label{sub:XP}
\para{Multi-dimensional Parallelism.}
 Large language models (LLMs) are typically composed of multiple transformer layers, each incorporating operators such as layer normalization (\texttt{LayerNorm}), general matrix multiplication (\texttt{GEMM}), 
etc. 
These operators form a Directed Acyclic Graph (DAG), and in the LLM training workflow,  
the data flow the corresponding DAG during the forward/backward propagation passes. 
To cope with the increasing model sizes and resource demands, today's 
training jobs are predominantly distributed to a cluster of GPUs, and adopt 
hybrid parallelism strategies~\cite{lin2024nnscaler,megatron-lm,zheng2022alpa}, coupling data and model parallelism.

With data parallelism (DP)~\cite{sergeev2018horovod,ren2021zero}, the entire model is replicated across GPUs/nodes.
Each model replica processes a subset of training data, followed by gradient synchronization at the end of an iteration. 
DP remains an important dimension of parallelism, though today's models are far beyond the memory capacity of a single GPU card or node, necessitating the simultaneous use of model parallelism.

The next ``outer loop'' in model parallelism is inter-layer parallelism, usually delivered by pipeline parallelism (PP)\cite{narayanan2019pipedream,huang2019gpipe,megatron-lm}, which splits model layers into different stages and assigns them to different GPUs/nodes. 
Each stage processes its assigned layers in a forward pass (1F) and backward pass (1B), known as the 1F1B pattern.

As both model sizes and sequence lengths grow, even a single layer does not fit in a GPU, requiring the exploitation of \textit{intra-layer parallelism}.
The latter in turn manifests in multiple dimensions, including tensor parallelism (TP)\cite{megatron-lm}, sequence parallelism (SP)\cite{korthikanti2023reducing}, context parallelism (CP)\cite{liu2023ring}, and expert parallelism (EP)\cite{he2021fastmoe,xue2024moeinfinity}, 
all dividing the computation of individual layers across multiple GPUs.

Among them, TP distributes a layer's computation across GPUs, by dividing its tensor data along one dimension, such as partitioning \texttt{GEMM} along the hidden size. 
Collective communication operators, like \texttt{AllGather} (\texttt{AG}) and \texttt{ReduceScatter} (\texttt{RS}), are then employed to aggregate intermediate output. 
Motivated by the ever-increasing context sequence sizes, SP complements TP by further distributing remaining operations like \texttt{LayerNorm} and \texttt{Dropout} across GPUs, by partitioning the input sequences. For very long sequences (e.g., 16K), context parallelism (CP) can be activated. It partitions and processes in parallel the token sequence, invoking \texttt{Allgather} communication to collect the corresponding key/value tensors from neighboring GPUs. 

Finally, for models using sparse architectures like Mixture of Experts (MoE)\cite{liu2024deepseek}, expert parallelism (EP)\cite{he2021fastmoe,xue2024moeinfinity} applies to multi-layer perceptron (MLP) layers. 
Here, different GPUs manage distinct parts of the MLP (referred to as \textit{experts}), again distributing tokens via \texttt{All-to-All} communication. 

Except for SP (which is always tied to TP) and EP (which is within DP), the above forms of parallelism in LLM training are orthogonal to each other and are often used in a hybrid manner.
State-of-the-art training frameworks like the prevalent NVIDIA Megatron-LM~\cite{megatron-lm} have already incorporated them to be used in a compounding manner. 
Here the type of parallelism often doubles as the parameter specifying the corresponding parallelism group size. 
For example, a Megatron-LM execution with the distribution setting \texttt{[TP=SP=8, CP=4, PP=8, EP=DP=4]} would run on a total of 1024 GPU cards. 
With common node settings installing 8 GPU cards per node, the above setting would perform TP and SP within each node, allowing their large amounts of inter-GPU communication to exploit the fast local-node network connection, while scaling out with CP, PP, and EP to additional nodes. 

\begin{figure}[!t]
    \centering
    \includegraphics[width=0.46\textwidth]{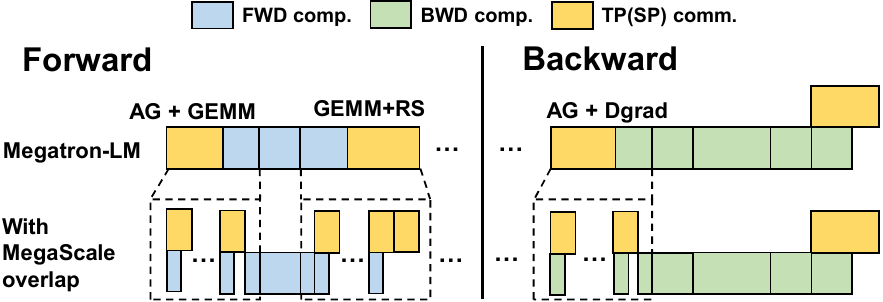}
    \caption{Sample result operator overlapping schedule by methods proposed in MegaScale~\cite{jiang2024megascale}, captured using the NVIDIA nsight profiling tool~\cite{Nsight}, in comparison to the execution follow achieved by Megatron-LM (top)
    }
    \label{fig:intra-batch-nsight}
\end{figure}

\para{Collective Communication.}
Partitioning and distributing computation and data inevitably incur communication, punctuating the computation phases with communication operators. 
Among them, \textit{collective communication} activities bring particular challenges to efficiency and scalability.

In the context of distributed LLM training, collective communication is associated with intra-layer parallelism: \textit{inter-layer parallelism} (PP) mostly passes data between neighboring stages.
For instance, TP involves data aggregation 
using \texttt{AllReduce}, which is transformed to  
\texttt{AllGather} and \texttt{ReduceScatter} when SP is incorporated.
The addition of CP further adds more \texttt{AllGather} operators to collect key/value tensors.
Finally, MoE introduces new communication patterns, with tokens shuffled by router gates between GPU experts using \texttt{All-to-All} operators.

Collective communication, which goes through the slower interconnection links and acts also as a global barrier, is known to be expensive and less scalable. 
Recent research, such as ring-attention~\cite{liu2023ring}, targets communication optimization such as replacing \texttt{AllGather} with more scalable \texttt{Send/Recv} operations.
However, as the model and sequence sizes keep growing, collective communication remains challenging for performance/cost optimization.

\subsection{Training with Input Batches}
\label{sub:batching}
The discussion above describes a “spatial” view of distributed
LLM training, while next, we give a “temporal” view.

The LLM pre-training process, which often takes days and weeks, iterates over trillions of input training tokens. 
They are firstly organized into \textit{global batches} and fed into the GPUs.
The GPUs hosting one copy of the model (with model parallelism) collectively process a \textit{micro-batch} at a time, 
walking them through the model and accumulating gradients.
When a global batch is completed, the gradients are synchronized across devices and used to update the model.

Processing a micro-batch involves one \textit{forward pass} followed by one \textit{backward pass}. 
From the perspective of a single GPU, at any given moment, unless idle, it is performing either one forward pass or one backward pass.

Processing one micro-batch after another, its long computation can be viewed as a virtual execution wave alternating between forward and backward passes, like one $sin$ curve or a DNA strand.
Within each pass, the strand is made of operators (like nucleotide bases) as the basic unit of GPU scheduling, performing computation- or communication-intensive activities.

\subsection{Existing Work on Intra-batch Scheduling}
\label{subsec:exist_intra}
Computation-communication overlapping is a widely adopted optimization in model parallelism, such as TP and PP.
Existing approaches~\cite{jangda2022breaking,wang2022overlap,chen2024optimizing,jiang2024megascale,li2023automated,chen2024centauri}  break down the computation and communication into fine-grained tasks to achieve efficient interleaving. 
For example, as shown in Figure~\ref{fig:intra-batch-nsight}, MegaScale~\cite{jiang2024megascale} decomposes large \texttt{AllGather} operators and subsequent \texttt{GEMM} or \texttt{Dgrad} (Backward Data Gradient)that consume the data gathered by the former ones into fine-grained \texttt{Send/Recv} operators and matrix tiles. Then, within a single micro-batch, MegaScale overlaps the partitioned communication and computation operators when there is no data dependence between them. 

Similarly, Megatron-LM~\cite{megatron-lm} and Ring-attention~\cite{liu2023ring} replace \texttt{AllGather} in CP with \texttt{Send/Recv} to overlap tiled attention computation.
Additionally, PP optimizations~\cite{zhuang2023optimizing,li2023fold3d} have focused on overlapping pipeline communication, such as tensor transmission, with computation, effectively reducing pipeline bubbles and improving efficiency.

However, the above optimizations are still constrained by the inherent sequential nature of single micro-batch processing, where the computation and communication operators cannot proceed in parallel due to data dependence.
We implemented the MegaScale overlapping method in Megatron-LM and observed that they only manage to overlap collective 
communication with adjacent \texttt{GEMM} operations, leaving a significant 73.9\% portion of computation/communication execution un-overlapped. This is illustrated by the sample MegaScale result schedule in Figure~\ref{fig:intra-batch-nsight}, where \texttt{AllGather} is not fully overlapped since \texttt{GEMM}'s lifespan is too short, while there exists a large portion of computation in the backward pass for which one cannot find enough communication operators to overlap with it. In summary, the single-\flow scheduling limits the opportunities for overlap between communication and computation.

\section{Motivation and Approach Overview}
\label{sec:understand}

\subsection{Megatron-LM Communication Profiling} 
\label{understand:subsec:comm}

\begin{figure}[!t]
 \centering
\includegraphics[width=0.42\textwidth]{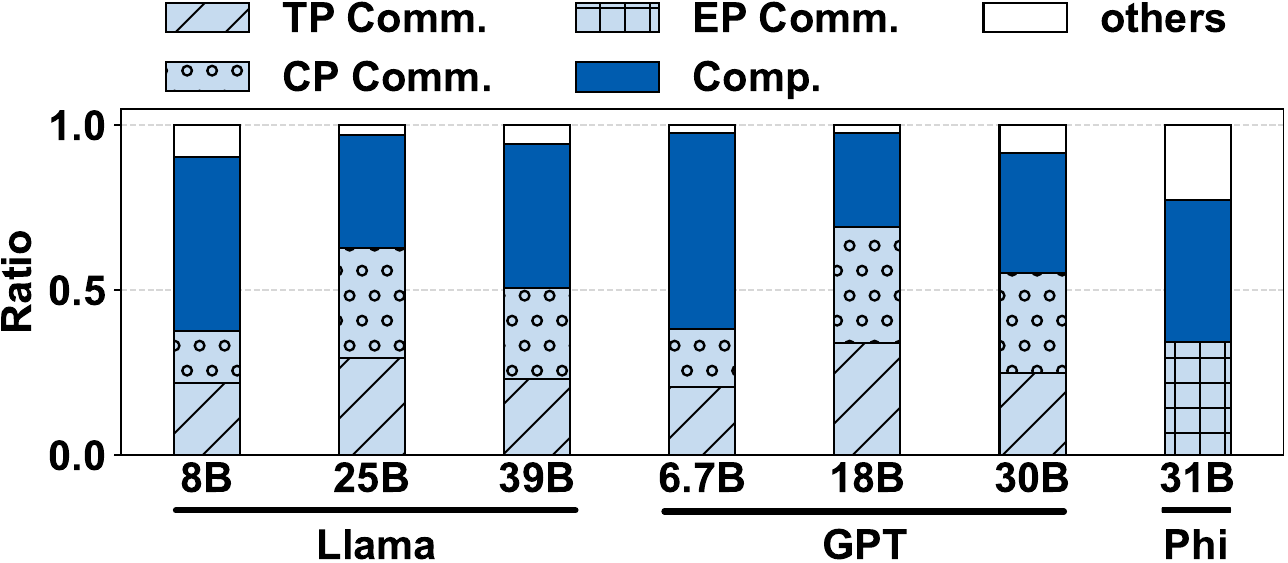}
 \caption{Sample execution time breakdown 
 in training different transformer-based models and parameter sizes}
 \label{fig:comp_ic_ratio}
\end{figure}

Our study begins by first analyzing the communication overhead during training, using the state-of-the-art Megatron-LM framework 
on a 64-card A40 GPU cluster.

Figure~\ref{fig:comp_ic_ratio} gives the distribution of the total pre-training time among computation and three types of 
communication operators for multiple transformer types and size combinations.
Communication has already become time-dominant at such a scale, especially with the larger models.
The major contributors are the collective  
communication operators brought by model parallelism, namely TP/SP, CP, EP. 
DP and PP, on the other hand, incur much lower communication volumes.
For example, with the Llama-39B model, TP- and CP-incurred 
communication occupies 55\% of the execution time; Phi-31B incurs around 34.3\% communication overhead under expert parallelism.

\begin{table}[!t]
    \centering
    \caption{Cross/local-node communication volume of Llama3.1-405B with hybrid parallelism.}
    \resizebox{0.47\textwidth}{!}{
        \begin{tabular}{|c|c|c|c|c|c|}
            \hline
            \multicolumn{1}{|c|}{\textbf{GPUs}} & \multicolumn{1}{c|}{\textbf{8192}} & \multicolumn{2}{c|}{\textbf{16384}} \\
            \hline
            \textbf{Parallelism} & DP:64,TP:8,PP:16 &
            DP:128,TP:8,PP:16  
            &DP:8,TP:8,PP:16,CP:16 \\
            \hline
            \textbf{Local-comm} & 0.98s (441 GB) & 0.3s (220GB) & 7.84s (3528GB) \\
            \hline
            \textbf{Cross-comm} & 0.17s (8.6 GB) & 0.15s (7.7GB) & 7.52s (376GB) \\
            \hline
        \end{tabular}
    }
    \label{tab:local_cross_volume}
\end{table}

It is common practice to keep large communication operators inside the machine. However, as more GPUs are needed to train larger models, the portion of cross-node communication will inevitably grow. Meanwhile, local communication speeds up when the inter-GPU bandwidth inside the machine is higher. The combination of the two has the potential to weaken the fact that intra-layer communication dominates. To understand this,  
we estimated the communication distribution between local- and cross-node communication for the LLM training on clusters using 10000+ GPUs. 
More specifically, we follow the distributed training configurations, and model architecture reported for Llama3.1 405B~\cite{dubey2024llama}, 
utilizing H100 GPU~\cite{h100-sheet} connected by NVLINK (900GB/s) and InfiniBand (3200Gbps) as network setup. 
We calculate the cross-node communication time costs based on communication volume and the bandwidth of NVLINK and InfiniBand. Table~\ref{tab:local_cross_volume} shows that the cross-node communication ratio over total communication time cost increases from 14.8\% to 33\% when moving from 8192 to 16384 nodes by simply scaling the DP group size. However, if CP is enabled, the cross-node communication ratio is further increased to 49.9\% as more cross-node Send/Recv is incurred. This indicates that 
communication remains a significant ratio during LLM training despite network development.

\begin{figure}[!t]
 \centering
 \includegraphics[width=0.50\textwidth]{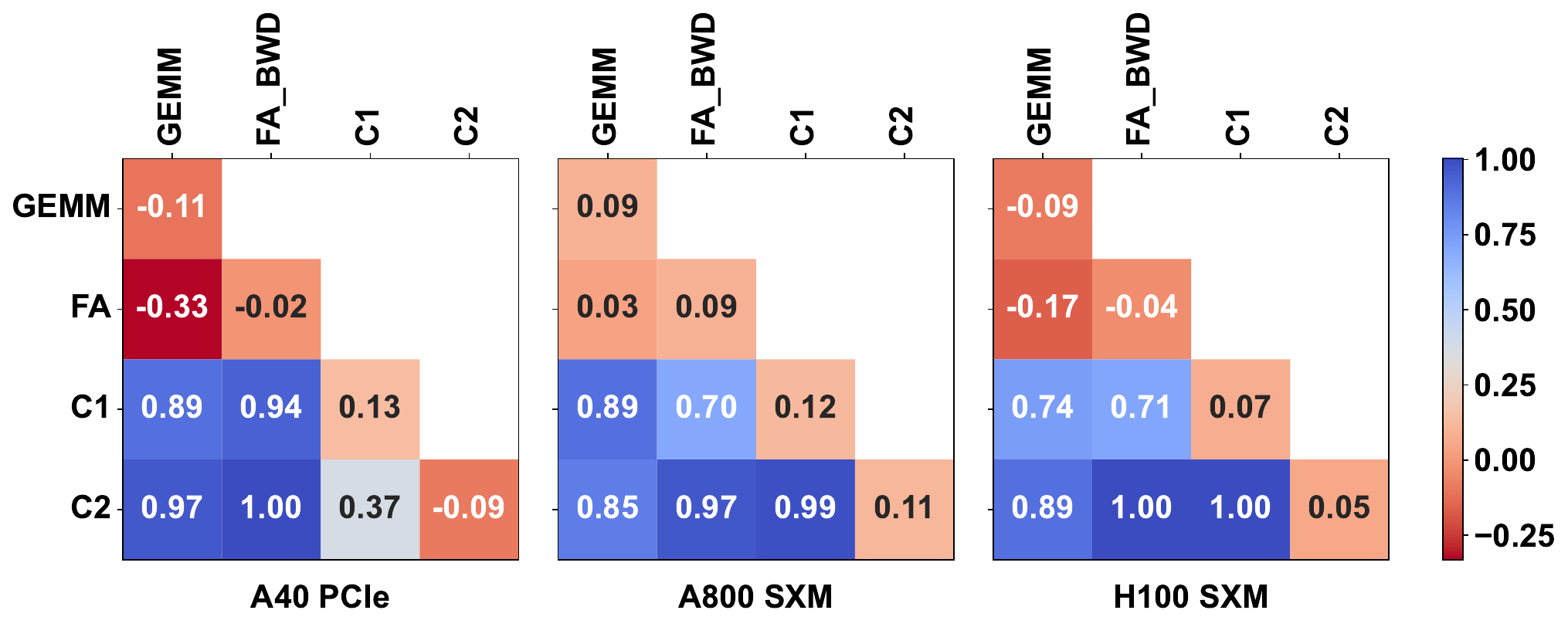}
 \caption{
 The overlap effectiveness is achieved through operator overlap. C1 represents local-node AllGather, while C2 denotes cross-node All-to-All. All operators, except for C2, are derived from the Llama 70B. 
 A complete pairwise table can be found in the 
 repository~\cite{H100-overlap-eff}
 , reporting the overlap effectiveness among 14 compute and 10 communication operators.
 }
 \vspace{-5pt}
 \label{fig:effectiveness}
\end{figure}

\subsection{LLM Training Operator Overlap}
\label{sec:overlap}

Though, as mentioned earlier, existing work has eagerly enabled the overlap between computation and communication activities to hide the cost of the latter~\cite{chen2024centauri,chang2024flux,jiang2024megascale,wang2024domino,wang2023zero++}, there lacks systematic evaluation of the performance behavior in overlapping commonly used LLM training operators.

In this work, we performed extensive profiling to understand the performance impact when we co-schedule two operators on the same GPU, with complete pairwise benchmarking, across multiple GPU types. 
Here we show results from 4 representative operators in Figure~\ref{fig:effectiveness}, two computation-intensive and four communication-intensive:
\texttt{GEMM}, \texttt{FA} (Flash Attention~\cite{dao2022flashattention,dao2023flashattention}, a highly optimized attention implementation), \texttt{C1} (local-node \texttt{AllGather}), and \texttt{C2} (cross-node \texttt{All-to-All}).

We measure the operator-level overlap behavior using a \textit{Overlap Effectiveness Factor (OEF)}, defined as
\begin{equation}
    OEF_{i,j}= (T_{i}+T_{j}-P_{i,j})/min(T_{i},T_{j})
    \vspace{-6pt}
\end{equation}
where $T_{i}$ is the sequential execution time of a single operator $op_i$, while  
$P_{i,j}$ is the overlapped execution time of $op_i$ and $op_j$.
In other words, OEF measures how much the overlap could hide the shorter operator's execution.

Figure~\ref{fig:effectiveness} gives the pairwise OEF across three NVIDIA GPU platforms: A40 (PCIe connection among GPUs on the same node), as well as A800 SXM and H100 SXM (both with NVLink for local-node, inter-GPU communication).
The major takeaways are:
\begin{itemize}[leftmargin=*]
\vspace{-3pt}
    \item Computation-intensive operators (\texttt{GEMM},  \texttt{FA}
    do not overlap well.)
    \item The \texttt{FA_BWD} operator is particularly unforgiving when overlapped with \texttt{GEMM}, likely due to the interference from the latter that breaks \texttt{FA}'s carefully choreographed fine-granule interleaving between computation and memory I/O.
    \item Both types of communication operators overlap well with both types of computation ones, though the degree of communication hiding achieved varies.
    \item Neither \local nor \cross communication operators overlap well with themselves, but benefit from overlapping with each other due to simultaneous use of different network resources (NVlink and Infiniband, for example).
\end{itemize}

These results inspire us to build an inter-batch execution interleaving scheme with systematic, fine-granule overlap optimization performed at the operator level. 
This allows for future model size scaling by relaxing the TP scaling constraint, currently limited to 8 (only among GPUs within the same node) in production model training due to its large communication volume.

\subsection{\si Overview}
\label{sub:si_overview}

\sys performs systematic interleaving at the operator level to accommodate two concurrent \flows, \afl and \bfl, each processing one micro-batch, for maximizing GPU utilization.

\begin{figure}[!t]
 \centering
 \includegraphics[width=0.45\textwidth]{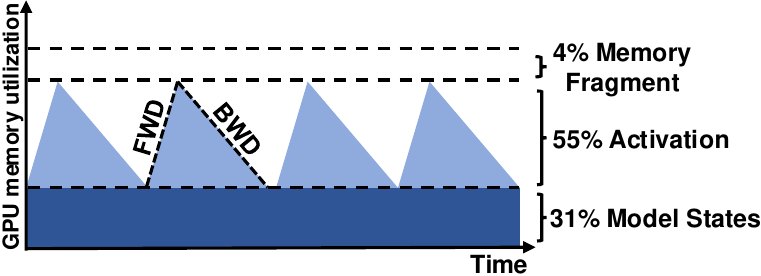}
 \caption{Sample memory allocation breakdown in training Llama-25B model, with 8192 sequence length and micro-batch size 1, on 64 A40 GPUs with parallelism strategy of DP=8 and TP=8.}
 \label{fig:footprint}
\end{figure}

Note that the common practice today, processing a single \flow, already maximizes the use of GPU memory by increasing the micro-batch size as much as possible for maximum training throughput. 
Figure~\ref{fig:footprint} illustrates the memory usage profile of a sample training run for a few micro-batches. 
Given the model size and parallel training parameters (Llama 25B, DP=8, TP=8), users typically bump up the micro-batch size (to 1), stopping right before the system runs out of memory.

The only way to squeeze in two \flows for \SI would be to introduce a time lag between them so that \afl's forward pass and \bfl's backward pass are co-scheduled, or vice versa, as shown in Figure~\ref{fig:overview}. 
The reason is that their execution has complementary 
memory consumption patterns: while the \flow running the forward pass steadily allocates memory space for activation data as it advances through layers, the one backward releases it at a similar pace. 
By fitting the two activation data ``triangles'' together, the total activation memory footprint stays close to the peak value from either single \flow.

\begin{figure}[!t]
 \centering
 \includegraphics[width=0.48\textwidth]{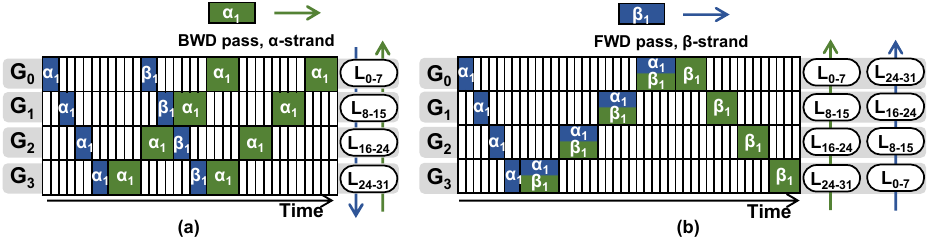}
 \caption{Pipeline scheduling with two strands: (a) V-shape 
 from 1F1B and (b) Bi-direction}
 \label{fig:v-shape}
 \vspace{-10pt}
\end{figure}

While the idea has been explored by the Wavelet approach~\cite{wang2021wavelet}, 
its proposed Tick-Tock scheduling, which was used in data parallelism and considers the overlap between the peaks and valleys of memory usage of the two strands. However, it falls short when applied to modern distributed LLM training setups for several reasons:
\begin{itemize}[leftmargin=*]
    \item \textit{Incompatibility with pipeline parallelism (PP).} The latter has become an essential mechanism for scaling LLM training, especially with large-scale training jobs, due to the relatively loosely coupled computation and lower overall communication volume (only across the boundary of pipeline stages). However, with PP, neighboring micro-batches forward and backward passes (the ``Tick'' and ``Tock'' waves in Wavelet) move in opposite directions. As shown in Figure~\ref{fig:v-shape}(a) at time 8, the forward (blue) pass of \bfl starts at GPU $G_0$, while the backward (green) pass of \afl starts at GPU $G_3$. The two \flows cross each other only once during the pass, leaving little opportunity for co-execution on the same GPU. 
    \item \textit{Model replication.} Wavelet co-schedules the two micro-batches of data training by replicating the model states (\ie, parameters, gradients, and optimizer states). Note that doing so might produce one solution to the PP scheduling problem by installing bi-directional pipelines, shown in Figure~\ref{fig:v-shape}(b). This allows the \afl and \bfl to move in the same direction, from $G_3$ to $G_0$. Unfortunately, with typical distributed training settings, model states occupy a considerable portion of GPU memory (over 30\% in the sample profiling result in Figure~\ref{fig:footprint}). 
    Storing two copies of the model would significantly forfeit the profit of micro-batch interleaving. 
    \item \textit{Coarse-granule interleaving.} In addition, Wavelet co-schedules the Tick and Tock waves to execute their original sequential workflow without intentional operator reorganization to exploit overlap opportunities provided by communication activities. 
\end{itemize}

\sys eliminates the above limitations with its \SI mechanism, which unlocks the cycle-efficient and memory-save
co-scheduling of two micro-batches processing on each GPU. 
To help picture \sys's working, imagine the double-helix DNA structure in a 2D space, whose two strands couple into a single stream of training computation, processing two micro-batches together.

The coupled ``DNA strands'', as a whole, get folded into a U-shape, with model layers doubling back across the GPUs 
participating in a PP pipeline. 
With this arrangement, we solve the PP incompatibility and the model replication problems listed above. 
As described in more detail in Section~\ref{sub:search}, the \bfl forward and \afl backward passes always move in the same direction in their lifetime, allowing their perfect co-execution from one device to the next in the PP scheme. 
Meanwhile, the U-shaped folding allows the model layers resident on each GPU  
to be shared by \afl and \bfl, simultaneously serving forward and backward passes with a single copy of model parameters.

When we zoom into the coupling of \bfl and \afl, the two \flows move up and down opposite each other, alternating between forward and backward passes. 
In each pass, the \flows rearrange their operators to find compatible operators that overlap well (\eg, computation with communication), creating a schedule anchored by these coupling points similar to the base pairs in DNA strands. 
Utilizing offline profiling results measuring the inter-operator overlap compatibility, \sys search for the efficient 
co-scheduling of the coupled \flows using dynamic programming, as to be discussed in Section~\ref{sub:search}.

As a result, \sys's \SI design enhances GPU utilization 
by enabling the training pathway to accommodate two neighboring micro-batches simultaneously, effectively hiding communication overhead in the critical path of LLM training, significantly boosting overall performance. 
Meanwhile, \SI operates below the existing levels of parallelism, seamlessly integrating with DP, TP, SP, CP, and EP. 
Our U-shaped folding technique further enables its compatibility with PP, where these forms of parallelism act as a basic GPU mesh for complete model training.

\section{\sys Design}
\label{sec:design}

\subsection{Model Folding} 
\label{sub:folding}

\begin{figure}[!t]
 \centering
 \includegraphics[width=0.4\textwidth]{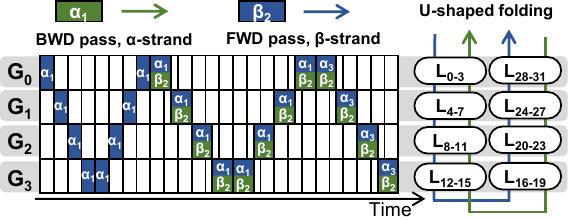}
 \caption{\SI with model folding (right side) in \sys and the resulted W-shaped scheduling (left side), in comparison to Figure~\ref{fig:v-shape}} 
 \label{fig:w-shape}
 \vspace{-10pt}
\end{figure}

\begin{figure*}[!t]
 \centering
 \includegraphics[width=\textwidth]{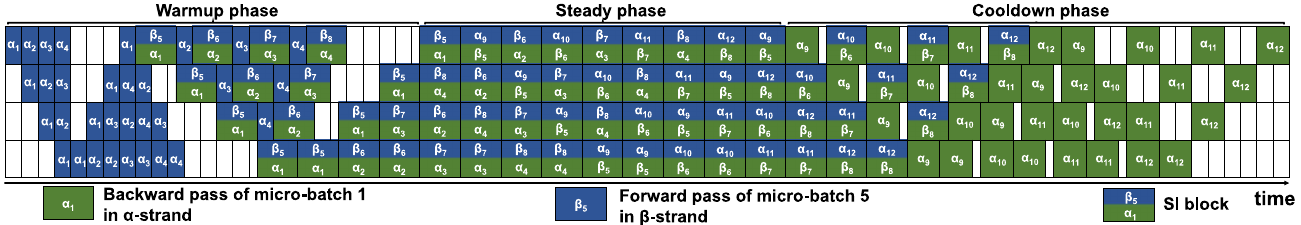}
 \caption{Sample W-shaped pipeline schedule in \sys 
 }
 \vspace{-5pt}
 \label{fig:w-pipeline}
\end{figure*}

Here, we describe \textit{model folding}. This key \sys technique enables \SI to work with pipeline parallelism, where we fold the original linear layout of model layers across GPUs to a U-shaped one.  
The right side of Figure~\ref{fig:w-shape} gives the layout of 
a 32-layer LLM after folding. 
The number of layers hosted per GPU remains at 8, but rather than hosting the eight consecutive layers $L_0$-$L_7$, $G_0$ now hosts two segments, $L_0$-$L_3$ plus $L_{28}$-$L_{31}$.

The difference is that now the flow of the two \flows across the GPUs are always heading the same way instead of opposite. 
The \afl backward (green) and \bfl forward (blue) passes start from $G_0$, reach $G_3$, and then return to $G_0$. 
In other words, as shown by the left side of Figure~\ref{fig:w-shape}, the familiar ``V'' shape of 1F1B becomes a ``W'' shape, where the forward and backward passes each make an identical ``V'' shape, allowing \afl and \bfl to perfectly overlap with each other in moving across the GPUs and their forward-backward passes to be co-scheduled on each GPU.

Compared with the model replication method discussed earlier (Figure~\ref{fig:v-shape}), \sys's model folding does not change the model parameter size per GPU. 
Therefore, with \SI, it accommodates both strands with the same copy of the model parameter, effectively processing two micro-batches on each GPU, while consuming almost the same GPU memory capacity as the state-of-the-art distributed training frameworks in processing one strand.

Figure~\ref{fig:w-shape} gives a simplified rendering of the double-\flow interleaved pass schedule, with the blue and green blocks possessing the same width (in time). 
It is well known that the backward passes are slower, with the green blocks almost twice as wide as the blue ones. 
When \SI co-schedules the \afl and \bfl operators on the same GPU, one might wonder if 
\afl's backward memory freeing could not keep up with \bfl forward memory requests and 
a space-induced dependency between the two \flows would be created. 
In reality, we found it to be a non-issue, and the difference would at most require a few hundred MBs of additional idle memory, 
as we overlap the forward and backward passes at the layer granularity instead of the whole model.

To give a more comprehensive picture, Figure~\ref{fig:w-pipeline} illustrates the complete W-shaped pipeline schedule with 12 micro-batches. 
It progresses through the same three phases as the classical 1F1B pipeline: 
\textit{warmup}, \textit{steady}, and \textit{cooldown}.
In the warmup phase, \sys populates the pipeline with $\alpha_{1}$-$\alpha_{4}$ (blue blocks) from \afl. 
Once they complete their forward passes, 
\sys injects $\beta_{5}$-$\beta_{8}$ from \bfl, beginning prefilling the pipeline via SI blocks (top blue and bottom green), with the backward passes in \afl and the forward passes in \bfl ``fused'' together.
Upon the issuing of the SI block 
$\left[ \genfrac{}{}{0pt}{}{\beta_5}{\alpha_1} \right]$, 
\sys moves into the 
steady phase. 
Here \sys launches $\alpha_{9}$-$\alpha_{12}$ in \afl and SI blocks occupy all GPUs. 
Finally, during the 
cooldown
phase, \sys runs out of forward tasks to create SI blocks and falls back to backward blocks (green) as the pipeline empties.

Both the original 1F1B and \sys's W-shaped schedule enable fully occupied GPUs during the steady phases. 
Also, in the warmup and cooldown phases, \sys only has the original single-strand execution (without SI). 
Therefore, \sys does not change the bubble ratio in the overall schedule, which remains at $\frac{p}{m-1}$. 
The performance benefit of \sys comes from enabling pipeline execution with SI in the first place (unlike Wavelet), then shortening the operator execution time by overlapping the two strands. 
In other words, each of the SI blocks 
$\left[ \genfrac{}{}{0pt}{}{\beta_i}{\alpha_j} \right]$
takes less time to complete than their sequential execution $\beta_{i}+\alpha_{j}$.

Also, as shown in the figure, the W-shaped pipeline requires two down-up trips for each micro-batch, doubling the \texttt{Send/Recv} communication volume from the original 1F1B scheduling. 
However, the share of communication volume attributed to such pipeline \texttt{Send/Recv} is minimal in common distributed LLM training and has little performance implication.

\subsection{Strand Coupling with Operator Pairing}
\label{sub:search}

Now, we zoom into a coupled forward+backward pass from the two strands (the combined $\alpha$ and $\beta$ blocks in Figure~\ref{fig:w-shape}). 
While the U-shaped model folding enables their co-execution, the operator-level overlap performed next brings \sys its major performance gain.

The gain comes from aggressively harvesting hardware parallelism whenever possible in co-executing the two \flows. 
As mentioned earlier, the individual operators in today's LLM training workflows, performing computation or communication, have been intensively optimized, enabling the overlap between these two activities~\cite{jiang2024megascale,chang2024flux}. 
As a direct consequence, the current operators have little chance to further overlap within each \flow due to data dependence that forces their sequential scheduling.

However, \SI opens up new space for overlapping \afl's communication operators with activities of \bfl, and vice versa, as there is no data dependence between the \afl and the \bfl. 
Here, in addition to the memory benefit of interleaving the forward and backward passes of the two \flows, there are two more performance advantages:
\begin{itemize}[leftmargin=*]
    \item The forward and backward passes have different operator layouts, leaving more space for co-scheduling computation with communication operators (as compared to forward-forward or backward-backward interleaving). 
    \item The backward pass tends to be more computation-intensive, in certain cases, providing more opportunities to hide the higher ratio of communication activities in the forward one.
\end{itemize}

\begin{figure}[!t]
 \centering
 \includegraphics[width=0.47\textwidth]{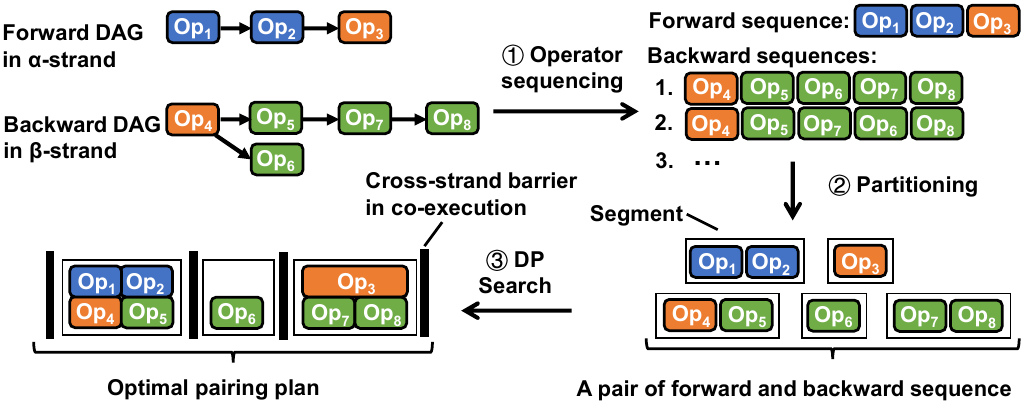}
 \caption{The workflow of strand coupling} 
 \label{fig:search}
 \vspace{-10pt}
\end{figure}

Rather than simply unleashing two micro-batches and letting the GPU do its best-effort co-scheduling (as with Wavelet~\cite{wang2021wavelet}), \sys carefully adopts a systematic and adaptive approach to intentionally align the two strands' execution at the operator level, to overlap computation and communication operators between two strands as possible. 
Figure~\ref{fig:search} illustrates its overall workflow: (1) generating all possible \textit{operator sequences} based on the appropriate DAG, (2) generating \textit{operator segments} by partitioning a pair of forward/backward operator sequences into contiguous pieces, and (3) using a dynamic programming algorithm to search for the optimal SI pairing scheme, which are administered by inserting barriers during the two strands' co-execution.
Below we discuss these steps in more detail.

\para{Operator Base Characterization.} 
Again, one could return to the DNA structure analogy, where the operators ``pair'' with appropriate peers in the opposite strand to form a ``bond'', in this case, co-execution that offers significant performance gain by utilizing heterogeneous GPU resources.  
Unlike the DNA strands, with four types of nucleobases, here conceptually, one could picture the transformer operators falling into only two types: computation or communication, as listed in Table~\ref{tab:op_base}.

Conceptually, the ``base pairing'' happens only between profitable base types: comp-comm or comm-comm, as seen in Section~\ref{sec:overlap}. 
In \sys design, to account for the intricate interleaving behavior between operators, we exhaustively perform pair-wise measurement of the operators listed above in Table~\ref{tab:op_base}. 
Such offline pre-profiling must be performed on a new hardware setup or when the training workflow is modified (such as updated operator implementation), and takes around 10-30 min.

\para{Operator Sequencing and Partitioning.} 
\sys has no control over the CUDA scheduling of individual operators. 
What it can do, however, is to throw barriers strategically across the execution of the \afl and the \bfl, forcing their synchronization. 
This way, communication operators could be co-executed with peers from the other \flow, hoping to maximize the chance of their cost being hidden by computation. In other words, the linear operator execution sequence of a single \flow is \textit{partitioned} into segments, which are then \textit{paired} across \flows, both via the \sys injected barriers.

\sys starts the \flow pairing process by constructing the 
DAG
for \textit{computing a single layer in the forward and backward passes}, 
composing 14 and 18 operators for one transformer layer if enabling tensor/sequence parallelism. For brevity, in the rest of our description, we often use terms like ``forward sequence'', though the DAG and resulting sequences are from one layer of the pass. Section~\ref{sub:discussion} will also discuss the current and future expansion of the pairing and search scope.
Based on these DAGs, as shown in Figure~\ref{fig:search}, \sys firstly generates the forward/backward candidate operator sequences by enumerating their topological orderings.

Naturally, the next step would be to partition each pair of candidate sequences into \textit{operator segments} (with barriers inserted in between), and co-schedule the segments across the two sequences (\flows).  Given a pair of candidate forward-backward sequences, positioning the cross-strand barriers effectively generates a \textit{candidate pairing plan}.

These two DAGs are simple with current transformer workflows, with a few operators that could move around in the result sequences. 
These include the backward weight gradient (\texttt{wgrad}, performing \texttt{GEMM}) and the \texttt{router} 
computation in MoE.
Still, the number of candidate sequences and their partitioning already result in a search space too large to be explored manually. 
Fortunately, the search for an optimal pairing plan can be formulated as a dynamic programming problem, as outlined below.

\begin{table}[!t]
    \caption{Major operator bases in transformer workflow}
    \label{tab:op_base}
    \centering
    \resizebox{0.47\textwidth}{!}{
        \scriptsize
        \begin{tabular}{ll|l}
            \toprule
            \multicolumn{2}{c|} {\bf Computation (comp)} & {\bf Communication (comm)} \\
            \midrule
            GEMM            & Fused BDA    & AllGather  \\
            FlashAttention  & Layernorm    & ReduceScatter \\
            Group-GEMM      & Router       & AlltoAll  \\
                            & Permute    & Send/Recv  \\
            \bottomrule
        \end{tabular}
    }
    \vspace{-10pt}
\end{table}

\para{Pairing by Dynamic Programming.} 
Given a pair of forward and backward pass candidate operator sequences, $S_f$ and $S_b$, each partitioned into $N_f$ and $N_b$ segments, \sys searches for the optimal operator pairing plan that yields the shortest total execution make-span $T_{opt}(N_f, N_b)$.

Intuitively, a pairing sub-solution for a pair of prefix sequences in the optimal solution, containing $i$ ($0\!\leq\!i\!<\!N_f$) and $j$ ($0\!\leq\!j\!<\!N_b$) operator segments from $S_f$ and $S_b$, must also be optimal: 

\begin{equation}
\begin{aligned}
T_{opt}\left(i,j\right)
= \min
\left\{
\begin{array}{c}
T_{opt}(i-1, j) + P(i,\emptyset),  \\
T_{opt}(i, j-1) + P(\emptyset, j), \\
T_{opt}(i-1, j-1) + P(i, j)
\end{array}
\right\}
\end{aligned}
\label{equal:eq2}
\end{equation}

Here $P(i,j)$ gives the overlapped execution time of overlapping the $i$th operator segment from the forward sequence with the $j$th operator segment from the backward one, calculated from the offline profiling results. 
$P(i,\emptyset)$ gives the $i$th operator segment's solo execution time, as it is scheduled to run alone.

\subsection{Discussion}
\label{sub:discussion}

\para{Adaptivity.}
As can be seen from the system design descriptions given in Section~\ref{sub:folding} and \ref{sub:search}, \sys's \SI scheme is quite general, independent of the actual distributed training framework composition and the underlying hardware. It could adapt to new training frameworks or hardware platforms by repeating its offline profiling and searching for an optimized strand interleaving plan. 

\begin{figure}[!tb]
 \centering
 \includegraphics[width=0.48\textwidth]{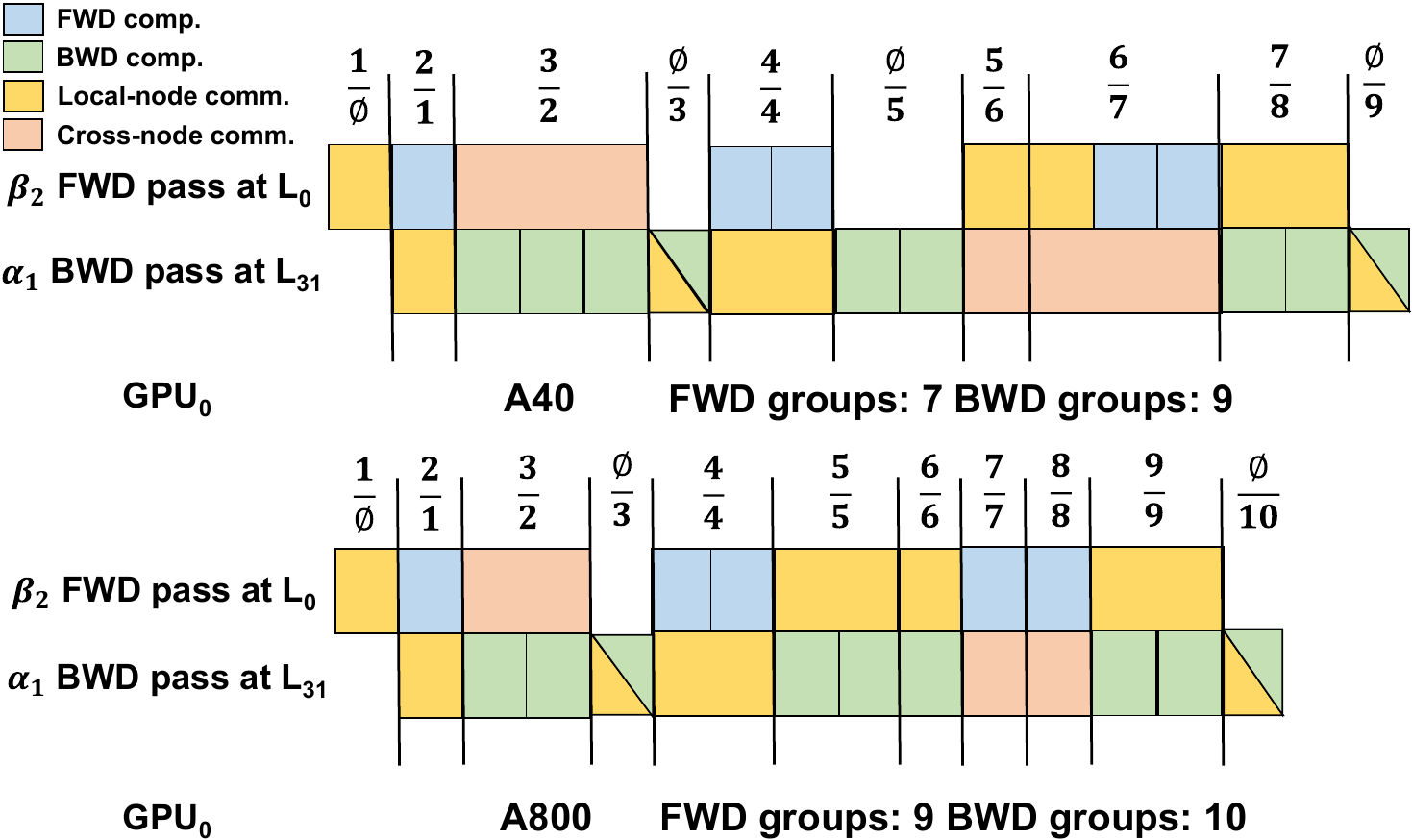}
 \caption{\SI pairing results on different hardware. Here the dual-color blocks note computation/communication operators in the backward pass that can be overlapped (no inter-dependency). Such ``self-overlapping'' operators within a \flow is treated as a single operator in \sys's pairing search.} 
 \label{fig:pairing_result}
 \vspace{-10pt}
\end{figure}

Figure~\ref{fig:pairing_result} demonstrates this with the search result of the same training workflow on two different platforms, on NVIDIA A40 (40GB) and A800 (80GB) clusters, respectively, showing quite different \SI plans. 
Though we do not have space here, it needs to be pointed out that the pairing schedule would also vary on the same hardware when one changes distributed training parameters (group size for TP, PP, SP, etc.), as the computation and communication operators' ``weight'' would change. 

Both AI model training frameworks and the underlying accelerator architectures are experiencing steady iterations of improvement in the years and even decades to come. 
Meanwhile, \sys's basic approach makes little assumption on their specific designs. 
More specifically, the model folding technique (Section~\ref{sub:folding}) applies to systems adopting neural network layers with backpropagation. 
The strand pairing technique (Section~\ref{sub:search}) is even more general, applicable to all training/inference workflows composed of operators that have complementary resource usage (such as tasks dominantly utilizing the GPU cores or the network connections).

\para{Scalability.} Finally, we highlight that aside from the operator characterization part (including the offline profiling), \SI remains transparent to model training users. If a framework execution with a set of multi-dimensional parallelism settings, the same settings can have \SI enabled without changing the parameters or requiring new \SI-specific parameters. Therefore, this makes \SI capable of optimizing the overall training throughput with very limited impact on user-visible software complexity. Testbed environments used in our evaluation (Section~\ref{sec:eval}), as limited by the hardware resources available to this research, can be viewed as a small patch of ``tiles'' that form stencil units to be replicated and connected by expanding in the PP, SP/CP, and EP dimensions for larger-scale training. Incorporating \SI does not affect a baseline framework's scaling out.

\section{Implementation Details}
\label{sec:impl}
We implemented \sys with around 5000 lines of Python code, on top of Megatron-LM~\cite{megatron-lm}, NVIDIA's popular distributed LLM training framework. 
The original transformer-based implementation in Megatron-LM consists of three modules: pre-processing, transformer, and post-processing. 
In \sys, we replaced the transformer module with the our \SI-enabled Transformer Block, to be connected with the other two Megatron-LM pre/post-processing modules for double-strand execution. 
All \sys components are based on \texttt{torch.nn.Module}, enabling users to create custom transformer models and take advantage of \SI using standard PyTorch APIs, requiring no user-level code changes.

At runtime, \sys processes the operator segment pairs sequentially according to the generated pairing plan. This sequential execution semantic is ensured by calling the \texttt{torch.cuda.synchronize} (\,  i.e., barrier) between two segments. 
Within the execution of one segment, operators are further categorized into three types: computation, local-node communication, and cross-node communication. We launch three CUDA streams and dispatch each type of operator to their dedicated stream for processing. 
Additionally, PyTorch allows different CUDA streams to independently allocate and release memory, which can lead to fragmentation and out-of-memory errors. To address this, \sys uses only the default CUDA stream for all memory allocation and deallocation operations.

In addition, we follow the existing practice in mitigating contention between computation and communication kernels~\cite{Liger}, by tuning the NCCL environment variables \texttt{NCCL\_\allowbreak NTHREADS} and \texttt{NCCL\_\allowbreak MAX\_\allowbreak NCHANNELS}.

\section{Evaluation}
\label{sec:eval}

\subsection{Experimental Setup}
\label{subsec:setup}
\para{Testbed.} 
We conducted experiments on three NVIDIA 
GPU clusters. 
The A40 cluster consists of 8 servers, connected through 100 Gbps InfiniBand, each with 8 A40 (48GB) GPU cards interconnected via PCIe4.0 at 32GB/s. 
The A800 cluster also has 8 servers with faster 4x200 Gbps InfiniBand interconnection, which has higher computing power and bandwidth than the A40 cluster. 
Each server has 8 A800 (80GB) GPUs , whose difference with the A100 cards lies in the weaker NVlink 4.0 local-node connection (400 GB/s).
The A100 cluster has 1 server with 8 A100(80GB) GPUs and NVLINk (600GB/s).
The H100 cluster has 4 servers connected with the fastest 8x400 Gbps InfiniBand, each with 8 H100 (80GB) GPUs and NVLINK (900GB/s).

The CUDA version used is 12.2. 
Due to the difficulty in requesting high-end GPU resources, the A800/H100 clusters were available to us for a very short period. 
We replicated our overall performance tests across testbeds and conducted the rest on the A40 one.  

\para{LLM/MoE Models.} 
As shown in Table~\ref{tab:llama_config},~\ref{tab:GPT_config},~\ref{tab:moe_config}, we test three mainstream open-source LLMs with nine different model sizes, ranging from 6.7 to 66 billion, including Llama~\cite{dubey2024llama}, GPT~\cite{brown2020language}, and Phi MoE~\cite{abdin2024phi}. 
Llama and GPT are two families of dense transformer-based models, while Phi is a sparse mixture-of-experts (MoE) LLM. 
Here, we set the top-$k$ value of MoE to 2, following its original setup, where the router gate forwards each token to top-2 relevant experts. 

In specific experiments, we adjust the number of model layers to fit in the cluster's aggregate GPU memory capacity based on the original model architecture. 
The performance results are expected to be very similar to the un-trimmed model size when the training scales out in the PP dimension, given the relatively small PP communication size.

\begin{table}[!t]
    \centering
    \caption{Llama settings used in our evaluation. We reduce the number of layers in Llama3.1-70B and Llama3.1-405B, resulting in models referred to as Llama-25B, Llama-39B, and Llama-66B, respectively.}
    \resizebox{0.47\textwidth}{!}{
        \begin{tabular}{|c|c|c|c|c|}
            \hline
            \textbf{Type} & \textbf{Hidden Size} & \textbf{Intermediate Size} & \textbf{\#Layers} & \textbf{Seq Len}  \\
            \hline
            \textbf{Llama-8B} & 4096 & 14336 & 32 & 8192,16384 \\
            \hline
            \textbf{Llama-25B} & 8192 & 28672 & 28 & 8192,16384 \\
            \hline
            \textbf{Llama-39B} & 16384 & 53248 & 12 & 8192,16384 \\
            \hline
            \textbf{Llama-66B} & 8192 & 28672 & 76 & 16384,32768 \\
            \hline
        \end{tabular}
    }
    \label{tab:llama_config}
    \vspace{-10pt}
\end{table}

\begin{table}[!t]
    \centering
    \caption{GPT settings in evaluation: the original GPT-6.7B and GPT-18B, plus GPT-30B (by reducing the number of layers from GPT-175B)}
    \resizebox{0.47\textwidth}{!}{
        \begin{tabular}{|c|c|c|c|c|}
            \hline
            \textbf{Type} & \textbf{Hidden Size} & \textbf{Intermediate Size} & \textbf{\#Layers} & \textbf{Seq Len}  \\
            \hline
            \textbf{GPT-6.7B} & 4096 & 16384 & 32 & 8192,16384 \\
            \hline
            \textbf{GPT-18B} & 6144 & 24576 & 40 & 8192,16384 \\
            \hline
            \textbf{GPT-30B} & 12288 & 49152 & 16 & 8192,16384 \\
            \hline
        \end{tabular}
    }
    \label{tab:GPT_config}
    \vspace{-10pt}
\end{table}

\begin{table}[!t]
    \centering
    \caption{MoE settings in evaluation: the original Phi-42B and its reduced-layer version Phi-31B }
    \resizebox{0.47\textwidth}{!}{
        \begin{tabular}{|c|c|c|c|c|c|c|c|}
            \hline
            \textbf{Type} & \textbf{Hidden Size} & \textbf{Intermediate Size} & \textbf{\#Layers} & \textbf{Seq Len}  & \textbf{\#Experts} \\
            \hline
            \textbf{phi-16B} & 4096 & 6400 & 12 & 3072 & 16 \\
            \hline
            \textbf{Phi-31B} & 4096 & 6400 & 24 & 3072 & 16 \\
            \hline
            \textbf{Phi-42B} & 4096 & 6400 & 32 & 3072 & 16 \\
            \hline            
        \end{tabular}
    }
    \label{tab:moe_config}
    \vspace{-10pt}
\end{table}

\para{Baseline systems.} 
We compare \sys with several baselines. 
Naturally, we compare with the vanilla \textbf{Megatron-LM}, the widely adopted distributed training framework supporting all forms of data/model parallelism discussed, including DP, TP/SP, CP, PP, and EP. 
Megatron-LM already includes communication overlap optimizations for CP, which overlap \texttt{Send/Recv} with attention computation.
The second baseline \textbf{Intra-batch} extends Megatron-LM by incorporating intra-batch communication overlapping, following the design of the industrial-strength MegaScale~\cite{jiang2024megascale}. 
It partitions \texttt{GEMM} operators and interleaves them with communication operators within a single batch for TP and SP. 

While these two baselines following single-\flow, 
the next one, \textbf{Wavelet+}, extends Wavelet~\cite{wang2021wavelet}, the only existing approach exploring micro-batch interleaving to our best knowledge.
As discussed earlier, 
it is applied only to DP and adopts model replication. 
For a fair comparison, we implement Wavelet's simple round-robin scheduling to interleave operators within \sys to form a pipelined Wavelet+. 
Note that Wavelet+ has \sys's model folding design, to remove its model replication and enable scheduling comparison using the same model sizes.

\para{Metrics.} 
Following existing practice~\cite{megatron-lm}, for Llama, GPT, and their variants, we report per-GPU training performance (TFLOPS/GPU) by dividing the job's total floating-point operation count on a single GPU by the total end-to-end execution time.
For the sparse Phi MoE models, we also follow related literature~\cite{xue2024moeinfinity} to measure by tokens/s, indicating
the generation speed. 
This is considering that 
some tokens may be dropped to balance computation among experts, the end-to-end TFLOPS cannot be calculated as accurately as with the dense models. Finally, \sys does not modify the semantics of distributed LLM training and thus does not affect the model convergence/accuracy.

\subsection{Overall Performance: A40 Cluster}
\label{subsec:overall}
First, we evaluate \sys with three tasks: dense model training, a dense model with long sequences and context parallelism (CP), and MoE model training. The global batch size is fixed at 8 million tokens for Llama/GPT, while the micro-batch size is adjusted to utilize GPU memory fully. 
Similarly, we set the global batch size to 1600 for Phi variants.

\subsubsection{Dense Models with Normal Sequence Length} 
\label{subsec:overall:dense:normallength}
\begin{figure}[!t]
 \centering
 \vspace{-10pt}
 \includegraphics[width=0.47\textwidth]{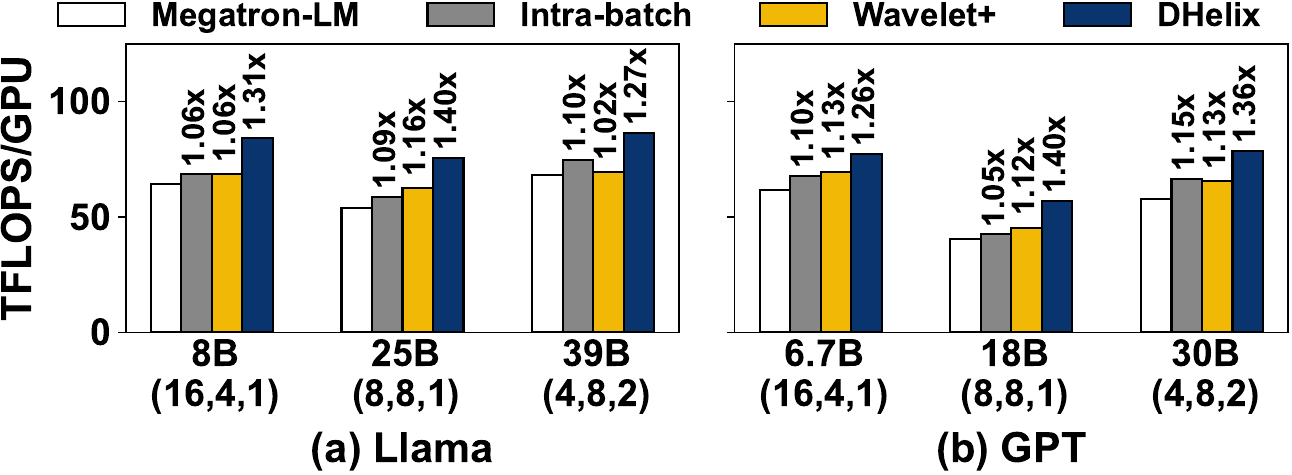}
 \vspace{-5pt}
 \caption{Overall performance w. Llama/GPT models. The tuple (x, y, z) under the bars gives DP, TP, and PP group sizes.} 
 \label{fig:task3d}
 \vspace{-10pt}
\end{figure}
Figure~\ref{fig:task3d} compares the performance of \sys against baselines on the Llama/GPT model with the default sequence length of 8192. 
The group sizes for tensor parallelism (TP) and pipeline parallelism (PP) are scaled according to the model size. 
For the Llama model, \sys consistently outperforms two baselines across different parallelism strategies and model sizes and achieves 27-40\%, 15-28\%, and 21-25\% improvement compared to Megatron-LM, Intra-batch, and Wavelet+, respectively.
GPT model training shows similar results, where \sys outperforms Megatron-LM, Intra-batch, and Wavelet+ by 26\%-40\%, 15\%-33\%, and 12\%-25\% improvement, respectively. 

Compared with the baseline Megatron-LM, both Intra-batch and Wavelet+ show much lower performance improvement (only 5-15\% and 6-16\%, respectively). Our profiling finds that Intra-batch hides only 26.1\% of the TP communication overhead by overlapping it with decomposed GEMM operators due to the sequential dependency within each micro-batch.
Wavelet+ performs micro-batch interleaving, but with its naive round-robin scheduling, it still only hides 39.9\% of the communication cost.
In contrast, \sys hides almost 83\%. 
Section~\ref{subsec:case} gives a detailed analysis.

\begin{figure}[!htb]
 \centering
 \vspace{-5pt}
 \includegraphics[width=0.47\textwidth]{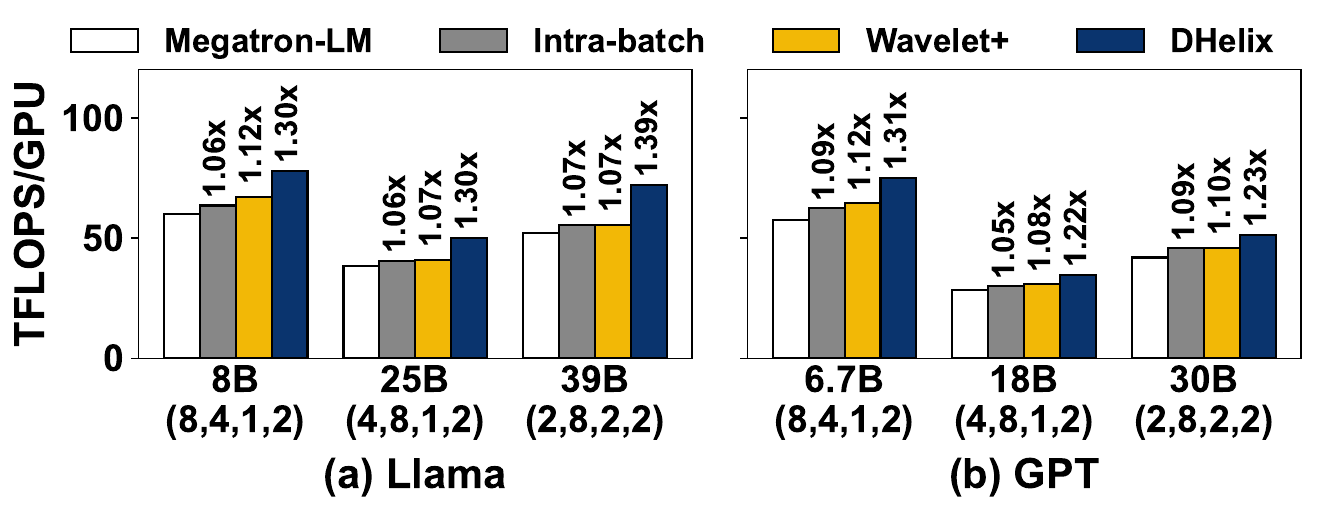}
 \vspace{-5pt}
 \caption{Llama/GPT training performance with long sequence. w in tuple (x, y, z, w) denotes CP group size.}
 \label{fig:task2}
 \vspace{-10pt}
\end{figure}
\subsubsection{Dense Models with Long Sequences} 
\label{subsub:long}
Considering the growing demand for long-context LLMs, we next evaluate Llama and GPT model training with long sequences, 
doubling the sequence length from 8192 to 16384 and adding context parallelism (CP) to all configurations. 
Note that adding CP leads to a reduction in the DP degree, as the total GPU number stays constant. 

Figure~\ref{fig:task2} shows that again \sys consistently outperforms Megatron-LM, Intra-batch, Wavelet+, with improvement ranging from 22\%-39\%, 13\%-30\%, 11\%-30\% respectively. 
Here, \sys's performance gains become even more pronounced because CP introduces additional intra-layer \texttt{Send/Recv} communication for exchanging KV blocks in attention computation, where \sys leverages its powerful and flexible pairing search to identify inter-strand overlap opportunities.

\subsubsection{MoE Models} 
Now we move to a sparse MoE model, Phi-31B, requiring EP to be enabled and
the DP group size to be divisible by the EP one. 
Following common practice~\cite{deepseek-hg,mixtral8-7B}, we set the EP group size to 8, with DP and PP group sizes set to 16 and 4 respectively. What's more, we set the capacity factor for each expert to 6. Note that the global parallelism group size is equal to 16$\times$4 as the EP group is a subset of the DP group.

\begin{table}[!htb]
    \centering
    \vspace{-5pt}
    \footnotesize
    \caption{The training performance table for Phi-31B with EP size of 8, DP size of 16, PP size of 4 and the capacity of the expert of 6 in A40 cluster.}
    \vspace{-5pt}
        \begin{tabular}{|c|c|c|c|}
            \hline
            \textbf{Metric} & \textbf{Megatron-LM} & \textbf{Wavelet+} & \textbf{\sys} \\
            \hline
            Token/s & 59941 &  62454 & 75386  \\
            \hline
            Improvement & - & 5\% & 27\%   \\
            \hline            
        \end{tabular}
    \label{tab:moe_metric}
    \vspace{-10pt}
\end{table}

Here, Wavelet+ can overlap communication with one short computation operation, bringing an improvement of 5\% over the Megatron-LM. 
\sys, on the other hand, delivers a 27\% gain due to its 
capability to effectively overlap the time-consuming \texttt{All-to-All} intra-layer communication brought by EP.

\subsection{Overall Performance on High-end Clusters}
\label{subsec:nvlink}
To evaluate the generality of our \sys's performance, 
We then extend our evaluation to the higher-end A800 cluster, with 80GB memory per GPU, NVLINK interconnects, and higher cross-node bandwidth. 
In our following experiments, we adapt model/sequence sizes and data/model parallelism settings accordingly to make full use of the resources. 
Note that our results here drop the ``Intra-batch'' bar, as it brings neglectable improvement with fast NVLink.
\begin{figure}[!t]
 \centering
 \vspace{-10pt}
\includegraphics[width=0.42\textwidth]{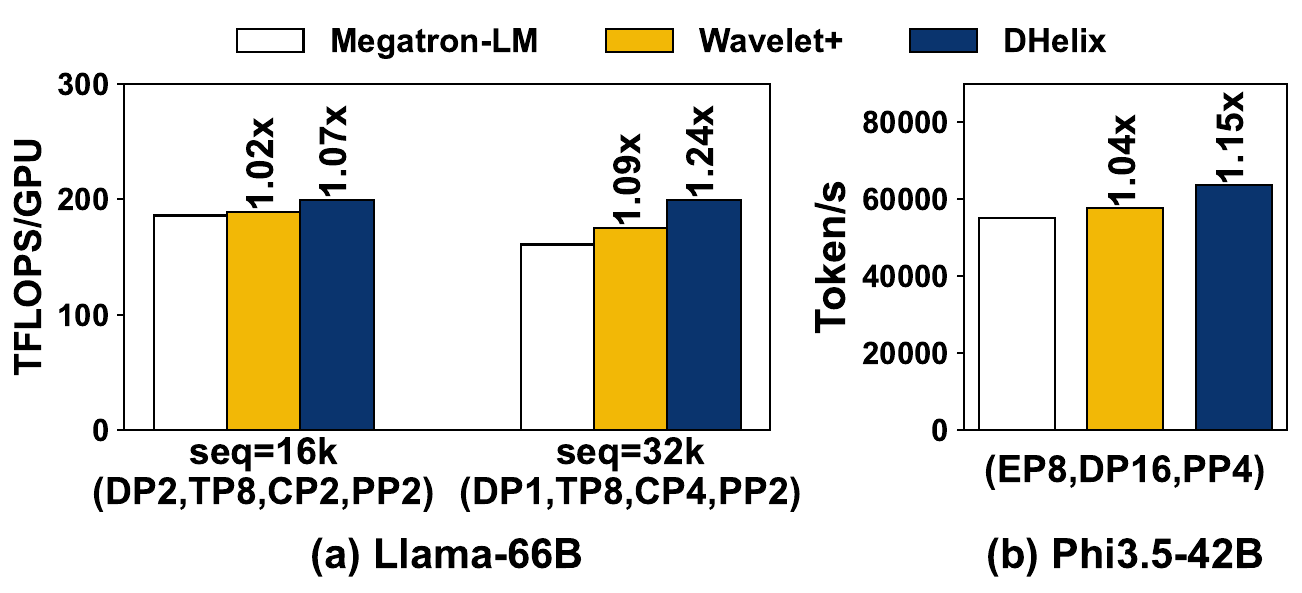}
\vspace{-5pt}
 \caption{A800 cluster training performance of (a) Llama 66B w. varied CP sizes, and (b) Phi-42B} 
 \label{fig:a800_moe_cp}
 \vspace{-10pt}
\end{figure}

\subsubsection{Dense Models with Long Sequences.} 
We can now raise the Llama size to 66B, the largest in our evaluation. The sequence lengths are also expanded to 16384 and 32768, with CP group size to 2 and 4 in accordance. 

Figure~\ref{fig:a800_moe_cp}-(a) presents the per-GPU training throughput. 
With high-speed local-node NVLINK, Megatron becomes a strong baseline, delivering impressive performance with approximately 186 TFLOPS (60\% MFU) and 160.9 TFLOPS (52\% MFU) in both cases. 
This is attributed to a reduction in local-node TP communication cost, which now accounts for only 10\% of the total training time, compared to 29\% on the A40 cluster, with CP size at 2. 
Additionally, Megatron-LM can overlap part of CP-related communication. 

Wavelet+ shows slight improvement over Megatron-LM due to its simplistic communication overlap strategies. 
In contrast, \sys still delivers a 7-24\% improvement over Megatron-LM (5-14\% over Wavelet+).
Also, increasing the CP size from 2 to 4 introduces more cross-node \texttt{Send/Recv}, hurting the performance of both Megatron-LM and Wavelet+. However, \sys effectively hides the increased cross-node communication, sustaining a throughput of 199.7 TFLOPS (64\% MFU).

\subsubsection{Larger MoE Model}

\label{sub:A800}
With twice the GPU memory capacity, the A800 cluster can accommodate a full Phi-42B model and we can set the capacity factor for each expert to 8, with performance reported 
in Figure~\ref{fig:a800_moe_cp}-(b).
Again, Wavelet+ achieves only 4\% improvement over Megatron-LM,
 while \sys delivers 15\%.  

\begin{figure}[!t]
 \centering
 \vspace{-10pt}
 \includegraphics[width=0.42\textwidth]{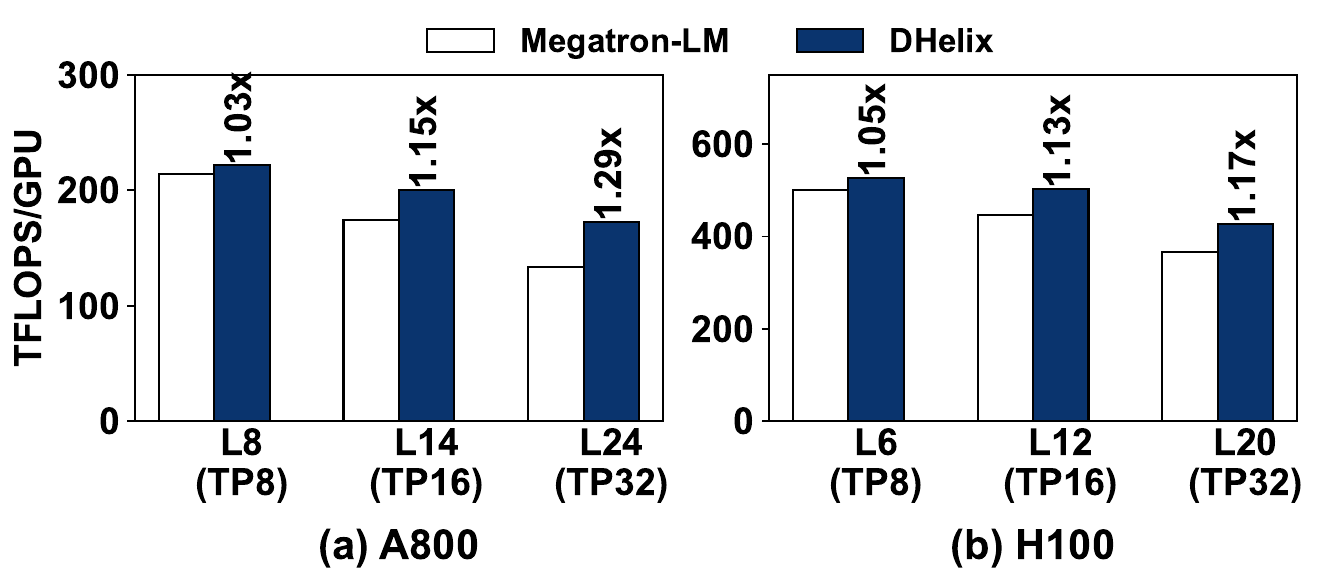}
 \vspace{-5pt}
\caption{Scaling model training with cross-node TP, w. transformer layer configuration from Llama3.1-405B~\cite{dubey2024llama}. 
For example, ``L8 (TP8)'' refers to a transformer model with 8 layers and a TP group size of 8.}
 \label{fig:tp_expand}
 \vspace{-15pt}
\end{figure}
\subsubsection{Increasing TP Group Sizes.}

Large-scale training efforts mostly limit the TP size to 8 (within a node), due to the prohibitive cost of TP-induced communication volume.
Recently, however, cross-node TP started to be examined~\cite{wang2024domino}, driven by growing model size and faster networking (e.g., Nvidia DGX and InfiniBand~\cite{ConnectX-8} offering up to 800GB/s aggregated bandwidth).
To this end, we obtain a short window of a 4-node H100 cluster to test \sys's potential in unlocking cross-node TP. 

Figure~\ref{fig:tp_expand} lists results with increasing TP size from 8 to 32. 
Note that higher TP size enables larger models (more layers),\footnote{Due to a runtime problem that we did not have chance to debug during our access to the clusters, we were not able to enable the same number of layers on H100 as on A800} trading off per-GPU throughput.

\sys sees lower profit over Megatron-LM on H100 (up to 17\%) than on A800 (up to 29\%) due to the former's much powerful interconnection (hence smaller visible communication cost).
However, on both GPU platforms it demonstrates the same advantage: the loss of TFLOPS from TP=8 to TP=32 is much slower while the model is over 3$\times$ larger.

To evaluate \sys's operator overlapping effectiveness on H100, we dive into communication overhead when training with cross-node TP. 
Although we have hidden 50\%-70\% of the visible communication cost, further optimization is possible. 
Table~\ref{tab:large_tp_bd} highlights two main factors limiting this overlap: 
(1) \textbf{kernel slowdown} -- running computation and communication kernels concurrently on multiple CUDA streams reduces performance by 20\%-30\%, limiting potential performance gain; 
(2) \textbf{launch interval} -- the small intervals between the launches of computation and communication kernels at the start of each paired segment pose an overhead amounting to 10\%-20\% of communication cost. 
A promising approach 
is to fuse these kernels into a single, monolithic kernel that manages GPU resources more precisely, as suggested in recent studies~\cite{li2022automatic,liu2023libra}.

\begin{table}[!htb]
    \centering
    \vspace{-5pt}
    \caption{Composition of communication cost measured in \sys, with different TP sizes on H100}
   \vspace{-5pt}
    \resizebox{0.47\textwidth}{!}{
        \begin{tabular}{|c|c|c|c|}
        \hline
        \textbf{TP size}     & \textbf{Hidden} & \textbf{Visible (slowdown)} & \textbf{Visible (launch interval)} \\ \hline
        \textbf{8}  & 57.87\%                        & 32.48\%                    & 9.65\%                   \\ \hline
        \textbf{16} & 66.73\%                        & 23.01\%                    & 10.26\%                  \\ \hline
        \textbf{32} & 51.72\%                        & 29.45\%                    & 18.83\%                  \\ \hline
        \end{tabular}
    }
    \label{tab:large_tp_bd}
        \vspace{-10pt}
\end{table}

\subsection{Improvement Breakdown}
\label{subsec:case}

\begin{figure}[!t]
 \centering
 \includegraphics[width=0.47\textwidth]{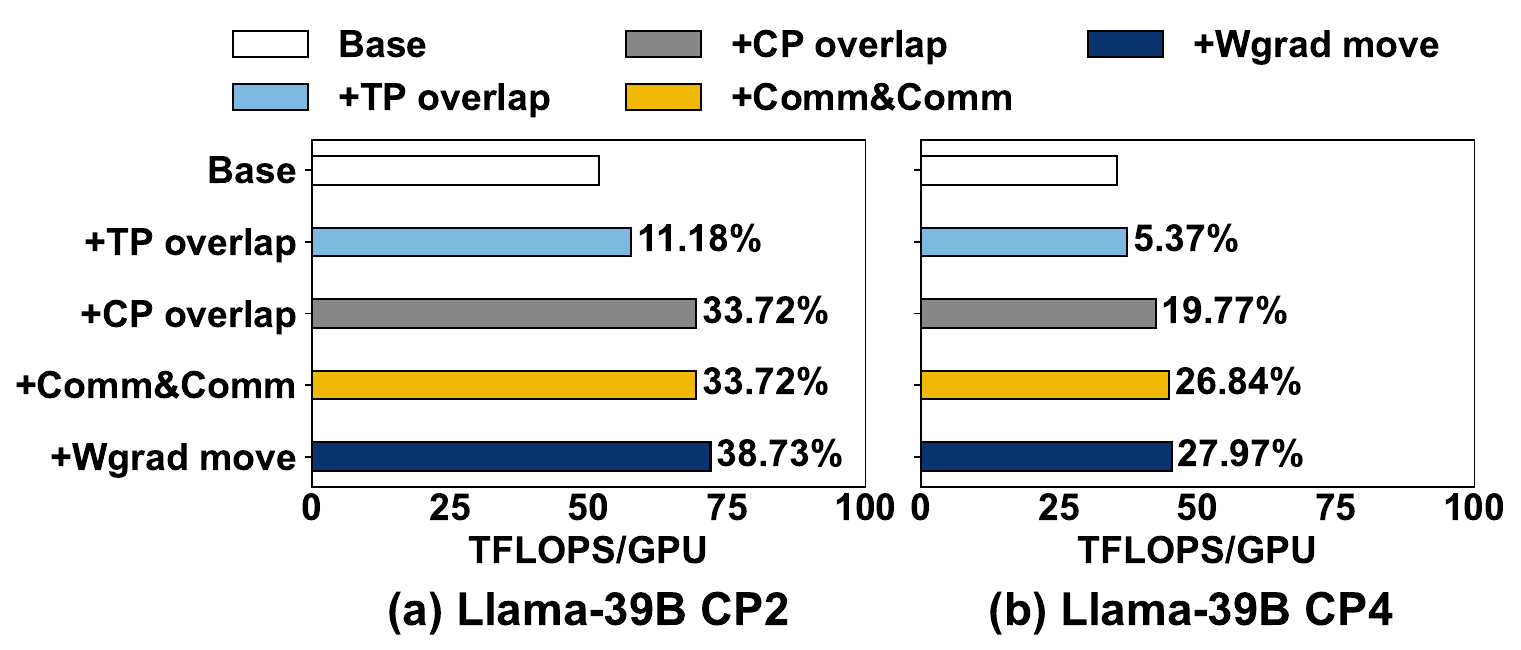}
 \vspace{-5pt}
 \caption{Breakdown of \sys improvement by incrementally overlapping communication types} 
 \label{fig:break.step}
 \vspace{-10pt}
\end{figure}

Next, we analyze the individual sources of \sys's performance gain over the Megatron-LM baseline (``Baseline'') 
of various communication overlap strategies we introduced to \sys. 
We are conducting breakdown experiments on the A40 cluster using the Llama-39B model in two scenarios: (a) CP at 2 and sequence length at 16384, and (b) CP at 4 and sequence length at 32768. 

Results are given in Figures~\ref{fig:break.step}-(a) and (b), respectively. Test (a) uses the same setup as Llama-39B in Figure~\ref{fig:task2}, while test (b) further evaluates the latency breakdown of \sys in a longer sequence length.

\para{TP Overlap.} Starting from Baseline, we first add TP overlap, which brings 11.18\% and 5.37\% improvements to Test (a) and (b), respectively. 
Typically, having higher communication costs is good news to \sys, as it sees a higher ``profit margin''. 
In Test (b), however, it turns into a curse as the communication overhead caused by the longer sequence length of 32768 is too high to overlap fully.
Faster interconnection could change the situation and allow \sys to gain more in these cases.

\para{CP Overlap.} We then turn on CP overlap, further hiding the CP-induced \texttt{Send/Recv} overhead, leading to another 22.5\% and 14.4\% improvement over the TP overlap bars, respectively.
This step contributes the most significant gain from \sys because CP-induced intra-layer communication accounts for 28\% and 53\% total training time in (a) and (b). 
Again, too much communication reduces the profit of CP overlap in the latter test.

\para{Communication and Communication Overlap.} 
The next addition of the overlap between local- and cross-node communication operators brings a further gain of about 7\% to Test (b), helping with its abundant communication activities. 
It does not improve Test (a), though, as most communication there has already been overlapped by computation.

\para{Wgrad Overlap.} 
Finally, we show the impact of operator reordering by enabling the \texttt{Wgrad} operator to move around as allowed by the DAG sorting. 
Test (a) yields an approximately 5\% improvement over the previous bar, which helps with ``overlap bubbles'' due to data dependency. 
In Test (b), there is insufficient computation to hide communication, and moving \texttt{Wgrad} only brings about 1\% performance gain.

\vspace{-5pt}
\subsection{Memory Space Utilization with \SI}
\label{subsec:max_model}

Finally, we verify \sys's memory optimization, which is crucial for supporting training jobs with model sizes as large as possible. 
We compare \sys's \SI scheme against Megatron-LM, the single-\flow baseline, and another straightforward but memory-unfriendly Bi-directional double-\flow implementation (illustrated in Figure~\ref{fig:v-shape}). 
The single-\flow represents the ideal maximum model size that GPU could support. 
In our experiment, we use the layer setup of the Llama-25B model in table~\ref{tab:llama_config} under the setup DP=4, TP=8, and PP=2, with micro-batch size set to 1. 
We increase the number of model layers to determine the maximum model parameter size supported by the system.

Our results find that \sys's maximum supported model size (39B) is close to Megatron-LM (40B), a mere 2.5\% model size reduction for a training throughput gain of 40\%. 
This slight decrease is due to \sys's scheduling certain forward operators in one strand being executed before the backward operators of the other, producing marginally higher peak memory usage. 
In contrast, without \SI's model folding solution, the bi-directional double-\flow implementation is forced to use model-state replication and, therefore, can accommodate models only up to 32B.

\section{Other Related Work}

\para{Communication overlapping optimization}
Communication overlapping techniques are widely used to mitigate communication bottlenecks in various parallelism strategies. For data parallelism(including ZeRO-family parallelism)~\cite{peng2019generic,jayarajan2019priority,hashemi2019tictac}, prior work has explored to better overlap the computation and communication by priority-based scheduling and tensor partition. Centauri~\cite{chen2024centauri} improves upon these works by graph-level scheduling.  
However, these existing communication overlap techniques are applied in single-\flow scheduling and are orthogonal to the double-\flow scheduling used by \sys.

\para{Other communication optimizations} 
Some research\cite{basu2024efficient,won2023tacos} focuses on optimizing collective communication operators based on hardware topology, improving network bandwidth utilization. Despite these improvements, intra-layer communication still constitutes a significant portion of end-to-end large language model training. MiCS~\cite{zhang2022mics} minimizes the communication size and reduces network traffic over slower links by leveraging a hierarchical model partition strategy. Other techniques\cite{bai2021hipress,wang2023espresso,song2023optimus,wang2023egeria} leverage the robustness of neural network training by compressing gradients or selectively transmitting part of the gradients during synchronization, thereby reducing the volume of communication. However, such communication optimization methods may negatively impact model training quality.

\para{GPU resource sharing.}
Existing research~\cite{xiao2018gandiva,lim2021zico,yu2019salus} has extensively explored GPU sharing techniques to fully harness GPU computing potential, focusing on temporal and spatial dimensions. Temporal GPU sharing involves software-based time-sharing mechanisms, where Gandiva~\cite{xiao2018gandiva} introduced a GPU time-slicing mechanism primarily to accelerate hyperparameter tuning jobs. This technique initiates job switching at iteration boundaries, reducing CPU-GPU communication overhead.

In contrast, spatial sharing techniques are equally crucial. Zico~\cite{lim2021zico} monitors memory usage patterns across training jobs, dynamically allocating and deallocating memory to ensure that reclaimed memory is globally available. MPS~\cite{MPS} partitions GPU memory statically for concurrent jobs, while TensorRT~\cite{tensorrt-infer} Inference Server enables simultaneous deep learning inference in parallel on a single GPU using GPU streams. Salus~\cite{yu2019salus} supports both temporal and spatial sharing by allowing rapid switching among DNN jobs and providing fine-grained memory abstraction. Additionally, pipeline-oriented optimizations in frameworks like Megatron-LM~\cite{narayanan2021efficient}, PipeDream, and Chimera~\cite{li2021chimera} execute computations for multiple microbatches simultaneously on the same set of GPU devices.

Compared to these approaches, \sys focuses on optimizing a single job by employing a sophisticated model to search for better scheduling strategies, thereby significantly enhancing GPU utilization.
\section{Conclusion}
We introduce \sys, a DNA-like abstraction for distributed LLM training that facilitates flexible overlaps between computation and communication of two co-executed micro-batches. 
\sys demonstrates that there is still considerable room for aggressively overlapping operators in the LLM training pipelines, and communication for cross-node tensor parallelism may be more manageable than it appears if we have a systematic mechanism to exploit the co-scheduling optimization space.


\bibliographystyle{plain}
\bibliography{ref}

\clearpage
\appendix

%
\label{appendix:case_study}

\begin{figure*}[!htb]
 \centering
\includegraphics[width=\textwidth]{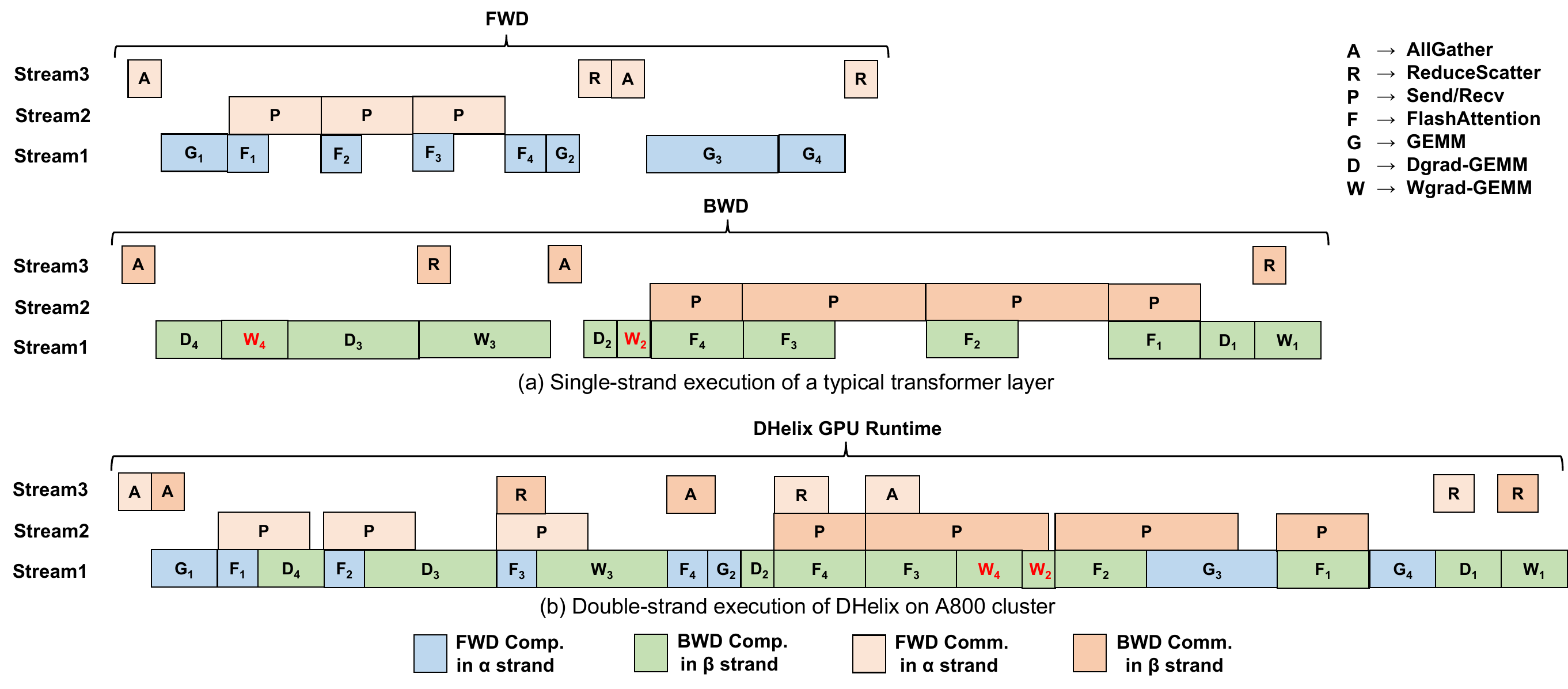}
 \caption{Visualization of the single-\flow and double-\flow CUDA schedule of SI on our A800 cluster, based on traces from the experiment shown in Figure~\ref{fig:a800_moe_cp} training Llama-66B (TP=8, CP=4).
 The top two rows ``(a) '' show the forward and backward passes executed without SI, respectively (where computation and communication have already been overlapped within each \flow.)
 The bottom row ``(b)'' shows the SI double-\flow execution, based on the optimal pairing found by dynamic programming. 
 For clarity, operators with negligible execution time (e.g., \texttt{LayerNorm}, \texttt{Dropout}) are omitted. 
 Here, \sys efficiently overlaps over 82\% of the total communication cost (\eg, TP-triggered local-node \texttt{AllGather} and \texttt{ReduceScatter} and CP-induced cross-node \texttt{Send/Recv} communication).
 The overall 
 speedup of \sys is 1.24$\times$, compared to single-\flow execution, as a result.
 } 
 \label{fig:cp_case}
\end{figure*}

\end{document}